# XANES Pb $L_{III}$ spectra of mixed-valence compound: Minium, $Pb_3O_4$

S.P. Gabuda*, S.G. Kozlova, S. B. Erenburg and N.V. Bausk

*Institute of Inorganic Chemistry of Siberian Department of Russian Academy of Sciences.*
*Novosobirsk 630090, Russian Federation*

___________________

*) Corresponding author: gabuda@casper.che.nsk.su

**Abstract.** Mixed-valence compound $Pb_3O_4$ (minium) has been studied using X-ray absorption near-edge structure (XANES) spectroscopy and DFT calculations. In spite of presence of two valence states of lead [Pb(II) and Pb(IV)], the XANES spectrum of studied system is corresponding to apparently unified, an intermediate valence state of Pb. On the other hand, the $^{207}Pb$ NMR spectra definitely show two different spectral bands corresponding to different $Pb^{2+}$ and $Pb^{4+}$ ions in $Pb_3O_4$ crystal structure. The explanation of this contradiction is related to the basics of XANES and NMR spectroscopy.

**Introduction**

Unusual feature of X-ray absorption near-edge structure (XANES) of Pb L(III) spectra of lead(II) oxides ($\alpha$-PbO and $\beta$-PbO) and of lead(IV) dioxide ($PbO_2$) is in that both are quite similar [1]. This fact is surprising in so far are different the electronic configurations: $5d^{10}6s^2$ for $Pb^{2+}$, and $5d^{10}$ for $Pb^{4+}$ ions. As a result, in $\beta$-$PbO_2$ is observed strong Pb L(III) absorption line connected with the allowed electronic dipolar transition $2p_{3/2} \rightarrow 6s_{1/2}$. Is clear, that similar absorption should not be observed for $Pb^{2+}$ where 6s state is occupied. Nevertheless, in both $\alpha$-PbO and $\beta$-PbO is clearly observed the Pb L(III) absorption at nearly the same energy as $2p_{3/2} \rightarrow 6s_{1/2}$ transition in $PbO_2$ [1]. Among the possible mechanisms of such absorption may be hypothetical occurrence of Pb=Pb localized chemical bonds, which can put the formal oxidation state in Pb(IV) for lead monoxides. Alternative model is related to the formation of the low-laying *s*-band (LUMO) built of virtual (unoccupied) energy levels of tunneling $6s^2$-electrons.

In the mixed-valence $Pb_3O_4$ (=2PbO·$PbO_2$) compound is expected a superposition of two independent XANES spectra if the first model is valid. In the other model, where delocalized bands are built of common AO of both Pb(II) and Pb(IV), most probably is expected the unified XANES spectrum reflecting the unified electronic structure of the bottom of the conductivity band of the crystal. Here we show, that the real XANES spectra of minium, $Pb_3O_4$, supports the last hypothesis. The discussed data may be useful in qualitative studies of chemical forms of elements in less-defined compounds where recently XANES spectroscopy is applied [2].

**Experimental**

$Pb_3O_4$ sample was prepared from PbO by oxidation under heating in air up to 500°C. Excess unreacted PbO was removed by dissolution in acetic acid. The X-ray powder data for $Pb_3O_4$ were the same as reported in the literature [3]. The chemical analysis of samples confirms the composition, with an impurity concentration that does not exceed 0.1%.

The XANES experiments were conducted at the EXAFS station of Synchrotron Radiation Center of Budker Institute of Nuclear Physics (Novosibirsk) based on storage ring of accelerator VEPP-3. In measurements, the storage ring operated at an energy of 2.00 GeV and a current of 50–100 mA. An ionization chamber filled with Ar/He was used as a monitoring detector. A mono block slit silicon single crystal ({111} plane) was used as a double crystal monochromator. The transmission spectra were recorded for powder samples pressed with an inert filler (cellulose).

**Results**

In our experiment was studied the Pb L(3)-edge fine structure of the X-ray absorption related to the electronic dipole transitions from the $2p_{3/2}$ core level to the unoccupied upper levels. The X-ray absorption L(3)-edge for metallic Pb is caused by the allowed transitions to the LUMO including $6d_{5/2}$ and $7s_{1/2}$ levels. The tabulated absorption edge energies is $Pb^0L(3) = 13.034$ keV [4]. With the aim of assigning the fine-structure components to particular electronic transitions, the transition energies $\Delta E$ were calculated for the $Pb^0$ atom and of $Pb^{2+}$, $Pb^{4+}$ ions using DFT method [5]. According to the data, the main peak is shifted to the higher energies in 8 eV for of bivalent, and in 19 eV tetravalent lead compounds. In addition, a fine structure appears at lower energies. In Figure 1 the calculated fine structure of the lead in Pb(II) and Pb(IV) valence states is represented by sticks with its heights relation 2:1. The calculated fine structure is confronted with the experimental XANES absorption spectrum of $Pb_3O_4$ (Figure 1, continuous line *a*) and with its second derivative where the fine structure of the spectrum is more pronounced (Figure 1, continuous line *b*).

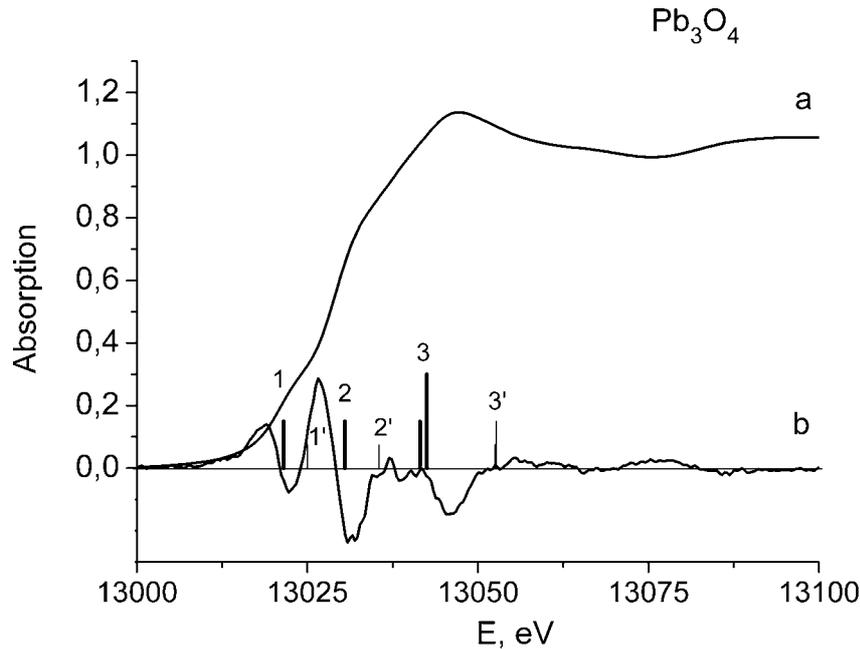

Fig. 1. The XANES Pb L(3) spectrum of $Pb_3O_4$ powder sample; *a* –absorption curve; *b*- its second derivative. Numbers above the sticks indicates the DFT-calculated components of fine structure, caused by transitions from the $Pb^{2+}$ basic $2p_{3/2}$ state to the excited states: 1 – $6s_{1/2}$; 2- $6p_{1/2}$; 3- $7s_{1/2}$ (left) and $6d_{3/2}$, and from the $Pb^{4+}$ basic $2p_{3/2}$ state to the excited states: 1' – $6s_{1/2}$; 2'- $6p_{1/2}$; 3'- $7s_{1/2}$ (left) and $6d_{3/2}$.

Generally, the excitation of the internal electrons in the X-ray absorption spectroscopy is accompanied by formation of short-living electronic hole. The typical time living of the excited state is shorter then $10^{-15}$ s, and hence, the natural width of the spectral absorption lines is of order 5 - 10 eV. This fact explains why the resolution of XANES spectra is generally pure. Nevertheless, the studied sample show pronounced fine structure of 3 lines visible both in absorption curve and, more clearly, in its second derivative. This fact is surprising in view of that calculated spectrum is consisting of 6 nearly uniform-spaced lines. Especially hard is to explain the absence of splitting of main peak at 13046 eV. This peak is related to the $Pb^{2+}$ and $Pb^{4+}$ electronic excitations of the basic-state $2p_{3/2}$ electrons to the $Pb^{2+}$ and $Pb^{4+}$ excited states $7s_{1/2}$ and $6d_{3/2}$ which are expected to be spaced in ~ 11 eV.

The possible explanation of absence of resolution of two strongest electronic transitions in $Pb_3O_4$ may be based on the fact that the excited electronic states of $Pb^{2+}$ and $Pb^{4+}$ in the crystal should be regarded as components of the lowest unoccupied molecular orbitals (LUMO) representing the bottom of delocalized conduction band. In such approach, the proper values of the energy of Hamiltonian operator of mixed-valence crystal cannot be related to subsequent excited electronic states of $Pb^{2+}$ and $Pb^{4+}$ ions, and do not be recorded separately within of XANES –experiment. It follows from such consideration that instead of individual valence states of lead in $Pb_3O_4$ XANES spectrum one may record only mean-weighted values of energies of electronic transitions:

$$<\Delta E[Pb(2p_{3/2})\rightarrow Pb(7s_{1/2})]> = 2/3\ \Delta E[Pb^{2+}(2p_{3/2})\rightarrow Pb^{2+}(7s_{1/2})] + 1/3\ \Delta E[Pb^{4+}(2p_{3/2})\rightarrow Pb^{4+}(7s_{1/2})]$$

where $\Delta E$ are DFT calculated energies. The same relation may be used for calculation of mean-weighted values of expected electronic transitions for $6s_{1/2}$ and $6p_{1/2}$ lines.

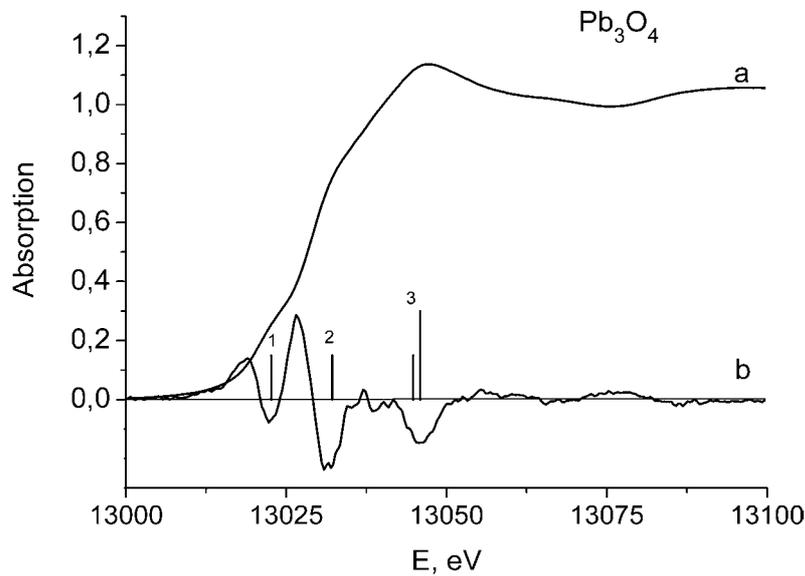

Fig. 2. Confrontation of XANES spectrum of $Pb_3O_4$ with the DFT-calculated and then averaged components: 1 – mean-weighted average of transitions from the lead basic $2p_{3/2}$ state to the $Pb^{2+}$ ($6s_{1/2}$) and to $Pb^{4+}$ ($6s_{1/2}$) states; 2 – the same average of $Pb^{2+}$ ($6p_{1/2}$) and of $Pb^{4+}$ ($6p_{1/2}$); 3 (left) – the average of $Pb^{2+}$ ($7s_{1/2}$) and of $Pb^{4+}$ ($7s_{1/2}$), and 3 (right) – the average of $Pb^{2+}$ ($6d_{3/2}$) and of $Pb^{4+}$ ($6d_{3/2}$).

In the Figure 2 is represented the calculated average XANES spectrum of $Pb_3O_4$. The comparison with experimental spectra show a good qualitative agreement with calculated one. This fact may be regarded as an evidence for the model of co-operative delocalized excited states of the bottom of the $Pb_3O_4$ conduction band involving both $Pb^{2+}$ and $Pb^{4+}$ atomic states.

**Discussion**

*Comparison with crystal structure data*

The room temperature crystal structure of $Pb_3O_4$ (Fig. 3) is tetragonal, space group $P4_2/mbc$, unit cell dimensions are $a$ = 8.82 Å; $c$ = 6.59 Å [3,6]. The crystal structure is comprised of chains of $[Pb^{IV}O_6]$ octahedra aligned along [001]. The space between $[Pb^{IV}O_6]$-chains is occupied by loose pairs of $Pb^{2+}$ ions, forming the columns along $4_2$ axis.

The nature of the $Pb^{2+}$ - $Pb^{2+}$ interaction in these pairs is still a matter of discussion, because the Pb-Pb distance in the pairs (3.799 Å) is markedly larger than the sum of $Pb^{2+}$ ionic radii (2.52 Å). It may be assumed, that such friable configuration of $Pb^{2+}$ ions is not stable, so the low-temperature structural transformations to the denser $Pb_3O_4$ lattice may be expected. In reality, $Pb_3O_4$ undergoes two low-temperature phase transitions (at 225 and 170 K), but with minor change of density: the unit cell dimensions are $a$ = $b$ = 8.8068 Å; $c$ = 6.5602 Å at ~ 170K in pseudotetragonal phase; and $a$ = 9.1152 Å; $b$ = 8.4696 Å; $c$ = 6.5646 Å at 30K in low-temperature orthorhombic structure [7].

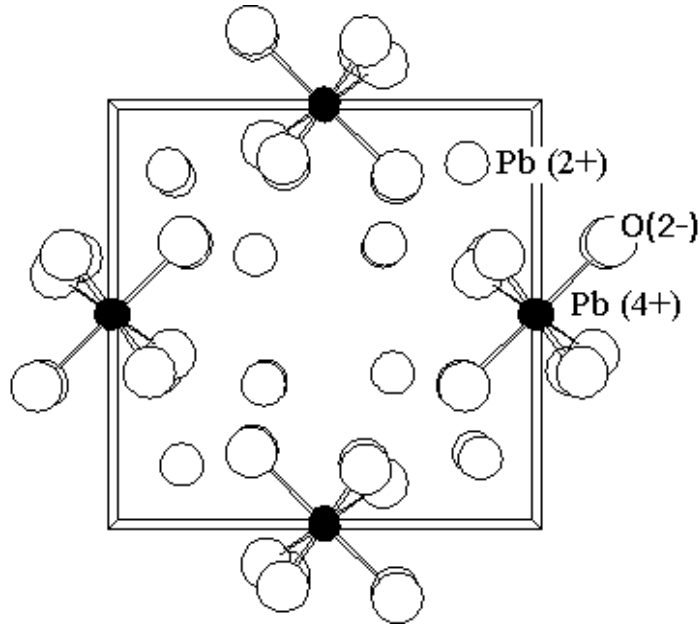

Fig. 3. The $Pb_3O_4$ crystal structure viewed along the [001]

It may be concluded from the $Pb_3O_4$ crystal structure data that $Pb^{II}$-$Pb^{II}$ interaction is strong enough for to split the $Pb^{II}(6s)$ state into the basic $Pb^{II}$ $\psi(6s)$, and the excited $Pb^{II}$ $\psi^*(6s)$ states. The basic $Pb^{II}$ $\psi(6s)$ state is populated by 2 electrons and is involved into the formation of highest occupied molecular orbital (HOMO) of the $Pb_3O_4$ valence band. The unoccupied excited states $Pb^{II}$ $\psi^*(6s)$ and $Pb^{IV}$ $\psi(6s)$ are supposed to form the above discussed LUMO of the $Pb_3O_4$ conduction band. Most probably, the known bright-red color of minium is due to excitation of $6s^2$ electrons from the basic state to the supposed LUMO of the $Pb_3O_4$ conduction band.

*Spectroscopy data*

From $Pb_3O_4$ vibrational spectroscopy study [8], the $Pb^{II}$-$Pb^{II}$ force field constants $f$(Pb-Pb)= 0.3 N/cm is of the same order as $Pb^{II}$-O force field constants: $f$(Pb-$O_1$) = 0.7; $f$(Pb-$O_2$) = 0.5 N/cm. Accounting the fact that Pb-O chemical bond is ionic in ~50%, it may be concluded that $Pb^{II}$-$Pb^{II}$ interaction is of the same order as contribution of exchange interaction in Pb-O bond.

*Comparison with $^{207}$Pb NMR spectroscopy data*

Vibration spectroscopy data correlates well with the $^{207}$Pb NMR spectroscopy data of $Pb_3O_4$ [9] showing that the chemical shift anisotropy is related mainly to the Pb-Pb interaction in clusters $[Pb_2]^{4+}$. $^{207}$Pb NMR spectra were recorded using a Bruker MSL-400 spectrometer ($B_o$ = 9.4 T; $\nu_o$ = 83.687 MHz) at room temperature. The pulse length was 0.7μs, and a relaxation delay of 0.5s was used. The sweep width was ~ 1 MHz. Typically, about $10^5$-$10^6$ free induction decays were accumulated. A typical, static $^{207}$Pb NMR spectrum of $Pb_3O_4$ is shown in Fig.4a. The spectrum consists of two separate bands, of which the areas are in the ratio of ~ 1:2. From this observation, we assign the narrow high-field line to $Pb^{4+}$, and the wide line to $Pb^{2+}$. The $^{207}$Pb NMR chemical-shift tensor of $Pb^{4+}$ in $Pb_3O_4$ is axial with values $\delta_{\parallel}$ = -1009±3; $\delta_{\perp}$ = -1132±3; $\delta_{iso}$ = -1091±3 ppm. The $Pb^{2+}$ chemical-shift tensor has principal values of $\delta_{11}$ = 1980±5; $\delta_{22}$ = 1540±5; $\delta_{33}$ = -1108±10; $\delta_{iso}$ = 804±10 ppm. The parallel component of the $Pb^{2+}$ line is less shielded than the perpendicular component.

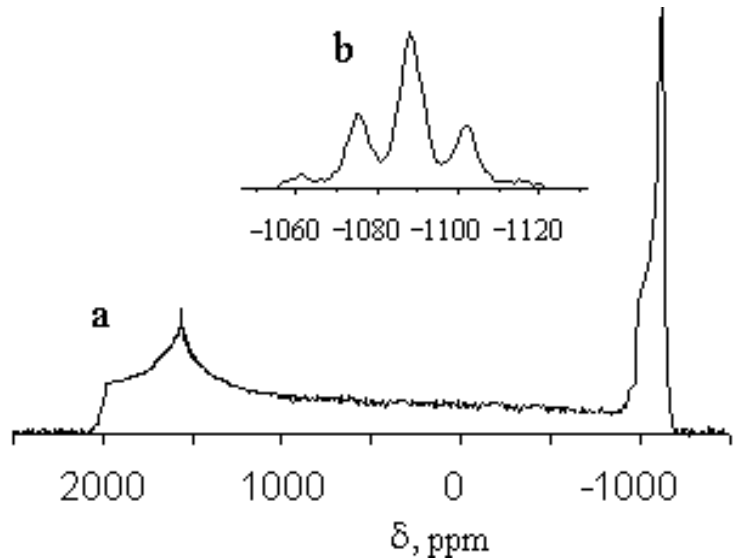

Fig. 4. $^{207}$Pb NMR spectra of Pb$_3$O$_4$ powder. Reference: Pb(CH$_3$)$_4$.
a) Static spectrum showing a wide Pb$^{2+}$ resonance (left), and the Pb$^{4+}$ resonance (right);
b) Centerband of the Pb$^{4+}$ MAS spectrum, showing the splitting due to Pb$^{II}$-Pb$^{IV}$ coupling.

At first glance it would seem that there is contradiction between the XANES and $^{207}$Pb NMR spectroscopy data of Pb$_3$O$_4$. This seeming contradiction is due to difference in the electronic basic-states encountered in two methods. The NMR chemical-shift tensor is connected with the mix of HOMO and LUMO states, and the HOMO states are different for Pb$^{4+}$ and Pb$^{2+}$ (5d$^{10}$ in the first, and 6s$^2$ in second). Otherwise, the Pb L (III) spectra are connected with electronic excitations from nearly the same 2p$_{3/2}$ energy-level of both Pb$^{II}$ and Pb$^{IV}$ in Pb$_3$O$_4$ to the same LUMO state.

*The nuclear Pb$^{II}$-Pb$^{IV}$ indirect coupling constant*

Some spectra were collected with the sample magic angle spinning (MAS), at ~5-12 kHz. Owing to the sample spinning the broadenings from the chemical shift anisotropy can be eliminated due to averagement. In Figure 4a is represented the typical $^{207}$Pb NMR MAS spectrum of Pb$^{4+}$ in Pb$_3$O$_4$ consisting of 5 equally spaced lines of relative amplitudes of ~1: 6: 11.5: 5: 0.8. Similar structure of $^{207}$Pb NMR MAS spectrum of Pb$^{2+}$ line was not observed because of its low signal-to-noise ratio.

The explanation of the observed MAS spectrum may be based on the model of nuclear spin-spin Pb$^{II}$-Pb$^{IV}$ indirect coupling [10]. In the Pb$_3$O$_4$ crystal structure [3], the nearest neighbours to the Pb$^{4+}$ ion are 8 Pb$^{2+}$ ions, the distances of which, D(Pb$^{4+}$ - Pb$^{2+}$), vary slightly. Half of them are involved in $Pb^{4+} - O_1 - Pb^{2+}$ bridges, which are shorter (3.62 A) than the other four (3.86 A). The MAS spectrum was simulated with a four-spin model of the interaction of the $Pb^{4+}$ nuclear spin with four other $^{207}Pb$ spins (abundance 21.11%) in the $Pb^{2+}$ sites by scalar coupling [10]. In such a model, the amplitudes of a calculated spectrum are in the ratio of 1: 5.2: 11.3: 5.2: 1, in close agreement with the observed relative intensities. The value of the coupling constant is $J(Pb^{IV}-Pb^{II}) = 2.3 \pm 0.1$ kHz.

**Conclusions.**

Above consideration suggests that the mixed-valence compound (minium, $Pb_3O_4$) is characterized by XANES spectrum which is connected with electronic excitations from the core $2p_{3/2}$ energy-level (of both $Pb^{II}$ and $Pb^{IV}$) to the delocalized LUMO forming the bottom of the conduction band of this crystal. The conclusion correlates well with the $^{207}Pb$ NMR MAS spectra of $Pb_3O_4$ showing the strong indirect exchange $Pb^{II} - Pb^{IV}$ interaction.

**Acknowledgements.** This work was supported by Division of chemistry and material sciences of RAS (Project No 15 of Program 4.1), and by RFBR (Grants 02-03-32816 and 02-03-32319).